\documentclass[a4paper,11pt]{article}
\usepackage{aaskaiid}
\usepackage{orcidlink}

\newcommand\farcs{\hbox{$.\!\!^{\prime\prime}$}}

\title{Pathfinder results for populations of planet-forming discs with SKAO }
\ShortTitle{Pathfinder results for planet-forming discs}

\author[1]{Jane S. Greaves\orcidlink{0000-0002-3133-413X}}
\ShortName{J. S. Greaves et al.} % shortened name list for header 
\author[2]{Anita M. S. Richards\orcidlink{0000-0002-6378-2594}}
\author[3]{Katie Hesterly\orcidlink{0000-0002-6378-2594}}
\author[2,4]{Emmanuel K. Bempong-Manful\orcidlink{0000-0002-1727-1224}}

\affiliation[1]{CHART, School of Physical, Chemical and Environmental Sciences, Cardiff University, 4 The Parade, Cardiff CF24 3AA, UK}
\emailAdd{greavesj1@cardiff.ac.uk}
\affiliation[2]{Jodrell Bank Centre for Astrophysics, Department of Physics and Astronomy, The University of Manchester, Manchester M13 9PL, UK}
\emailAdd{a.m.s.richards@manchester.ac.uk}
\affiliation[3]{SKAO, Block A, 2nd Floor, Observatory, Cape Town 7925, South Africa}
\affiliation[4]{School of Physics, University of Bristol, Tyndall Avenue, Bristol BS8 1TL, UK}

\abstract{Imaging of few-millimetre-sized dust grains (pebbles, on the construction path to planets) in protoplanetary discs has begun. The Planet-Earth Building-Blocks Legacy e-MERLIN Survey is a unique pathfinder for SKAO, imaging a sample of disc-host stars around 4.4 cm wavelength at 55--120 mas resolution. PEBBLeS has surveyed 10 flux-ordered fields of the northern sky, and the L1448C-N and HL Tau systems are presented as examples. The prospects for extending this science with SKAO are bright, including shorter integration times and better sensitivity to fainter outer discs. Simulations are presented of how a southern analogue to the HL Tau disc could look to SKA-Mid AA$^{\ast}$  in bands 5a and 5b. The initial lack of long baselines is suggested to pose a significant risk for imaging of planet-related features in protoplanetary discs. }

\begin{document}
\newcommand{\actaa}{Acta Astron.} % Acta Astronomica
\newcommand{\araa}{ARA\&A} % Annual Review of Astron and Astrophys
\newcommand{\aar}{A\&ARv} % Astrononmy \& Astrophysics Review
\newcommand{\aapr}{A\&ARv} % Astronomy\&Astrophysics Reviews
\newcommand{\ab}{Astrobiol.} % Astrobiology
\newcommand{\aj}{AJ} % Astronomical Journal
\newcommand{\apj}{ApJ} % Astrophysical Journal
\newcommand{\apjl}{ApJL} % Astrophysical Journal, Letters
\newcommand{\apjs}{ApJSS} % Astrophysical Journal, Supplement
\newcommand{\ao}{Appl. Opt.} % Applied Optics
\newcommand{\apss}{Astro. \& Space Sci.} % Astrophysics and Space Science
\newcommand{\aap}{A\&A} % Astronomy and Astrophysics
\newcommand{\aaps}{A\&AS.} % Astronomy and Astrophysics, Supplement
\newcommand{\baas}{Bull. Am. Astron. Soc.} % Bulletin of the AAS
\newcommand{\caa}{Chinese A\&A} % Chinese Astronomy and Astrophysics
\newcommand{\cjaa}{Chinese J. A\&A} % Chinese Journal of Astronomy and Astrophysics
\newcommand{\cqg}{Class. Quantum Gravity} % Classical and Quantum Gravity
\newcommand{\gal}{Galaxies} % Galaxies
\newcommand{\gca}{Geo. Cosmo. Acta} % Geochimica Cosmochimica Acta
\newcommand{\icarus}{Icarus} % Icarus
\newcommand{\jcap}{JCAP} % Journal of Cosmology and Astroparticle Physics
\newcommand{\jgr}{J. Geophys. Res.} % Journal of Geophysics Research
\newcommand{\jgrp}{J. Geophys. Res. Planets} % Journal of Geophysics Research: Planets
\newcommand{\jqsrt}{J. Quant. Spectrosc. Radiat. Transf.} % Journal of Quantitiative Spectroscopy and Radiative Transfer
\newcommand{\memsai}{Mem. SAIt} % Mem. Societa Astronomica Italiana
\newcommand{\mnras}{MNRAS} % Monthly Notices of the RAS
\newcommand{\nat}{Nature} % Nature
\newcommand{\nastro}{Nat. Astron.} % Nature Astronomy
\newcommand{\ncomms}{Nat. Commun.} % Nature Communications
\newcommand{\nphys}{Nat. Phys.} % Nature Physics
\newcommand{\na}{New Astron.} % New Astronomy
\newcommand{\nar}{New Astron. Rev.} % New Astronomy Review
\newcommand{\physrep}{Phys. Rep.} % Physics Reports
\newcommand{\pra}{Phys. Rev. A} % Physical Review A: General Physics
\newcommand{\prb}{Phys. Rev. B} % Physical Review B: Solid State
\newcommand{\prc}{Phys. Rev. C} % Physical Review C
\newcommand{\prd}{Phys. Rev. D} % Physical Review D
\newcommand{\pre}{Phys. Rev. E} % Physical Review E
\newcommand{\prx}{Phys. Rev. X} % Physical Review X
\newcommand{\prl}{Phys. Rev. Let.} % Physical Review Letters
\newcommand{\psj}{Planet. Sci. J.} % Planetary Science Journal
\newcommand{\planss}{Planet. Space Sci.} % Planetary Space Science
\newcommand{\pnas}{Proc. Natl Acad. Sci. USA} % Proceedings of the US National Academy of Sciences
\newcommand{\procspie}{Proc. SPIE} % Proceedings of the SPIE
\newcommand{\pasa}{PASA} % Publications of the Astron.  Soc. of Australia
\newcommand{\pasj}{PASJ} % Publications of the Astron.  Soc. of Japan 
\newcommand{\pasp}{PASP} % Publications of the Astron.  Soc. of the Pacific
\newcommand{\rmxaa}{RMXAA} % Revista Mexicana de Astronomia y Astrofisica
\newcommand{\sci}{Science} % Science
\newcommand{\sciadv}{Sci. Adv.} % Science Advances
\newcommand{\solphys}{Sol. Phys.} % Solar Physics
\newcommand{\sovast}{Soviet Ast.} % Soviet Astronomy
\newcommand{\ssr}{Space Sci. Rev.} % Space Science Reviews
\newcommand{\uni}{Universe} % Universe

\maketitle

\section{Introduction}

Radio interferometry is essential to understanding the growth of very large dust grains, or ‘pebbles’, in the discs around young stars. Pebble accretion may be a vital step to forming rocky planets, such as those in the solar system (e.g. \citealt{2025E&PSL.65019120M, 2025OJAp....8E.134P}). The black/grey-body emission of pebbles can be optically thick in the midplane – where planetary cores are presumed to form -- even to centimetre wavelengths. \citet{2019ApJ...877L..18Z} %Zhu et al. (2019)
calculate that a massive (Toomre parameter $Q = 1$, marginally unstable against collapse) disc could be opaque inside 10 AU radius even at 10 GHz ($\lambda$ 3 cm). Hence to look at the seeds of planets out to Saturn-like orbits, even longer wavelengths may be needed. \citet{Garufi01.2026.SKA} 
%\citet{2026aask.bookE..36G} 
discuss the potential with SKAO. 

Blackbody-like emission from grains is less efficient at emitted wavelengths that are (classically) larger than of-order the grain circumference \citep{2006ApJ...636.1114D}. Hence the use of long wavelengths can allow discovery of emission from the largest particles in the disc, the pebble population. The spectrum of grains that are large compared to the observing wavelength is simplistically $F_{\nu}\propto \nu^2$, where $F$ is the flux from grains measured at frequency $\nu$. However if scattering is included (e.g. \citealt{2017ApJ...850..115S}),
%Sierra et al. 2017),
an optically thick disc of grains can have a spectrum with an index that is more or less than 2 \citep{2019A&A...622A..75L, 2019ApJ...877L..18Z}.  This additional physics affects how the largest grain sizes present are inferred. 
%(Liu 2019, Zhu et al. 2019).

Young star systems exhibit multiple emission mechanisms at centimetre wavelengths -- for recent survey analyses,  \citet{2025OJAp....8E.134P, 2025ApJS..277...45C}.
%Caleb et al. (2025) and Chung et al. (2025).
Hence, a high angular resolution is needed to disentangle the disc signal from emission coming from the star and its jets. This can not be done reliably by spectral fitting of data collected over time, as electron emission processes often have strong temporal variation \citep{2015ApJ...801...91D}. %(Dzib et al. 2015).
High angular resolution allows the stellar sources and the orthogonal disc and jets to be spatially separated.

\section{Pathfinder science with e-MERLIN}

The pioneering study of TW Hya at 3.5 cm \citep{2005ApJ...626L.109W} %(Wilner et al. 2005)
gave the first indication of a pebble disc. Exploiting the upgrades of e-MERLIN taking place soon afterwards \citep{2016A&G....57c3.28G}, the Planet-Earth Building-Blocks Legacy e-MERLIN Survey (PEBBLeS) was approved in 2009 for a prototype study, and fully approved in 2017. Data acquisition finished in 2024 and data products will soon be made public, in conjunction with the survey definition and results paper (Greaves, Hesterly, Bempong-Manful, Chandler \& Richards, in prep., hereafter G26). 

PEBBLeS is an unbiased flux-ordered survey of young discs in the northern sky at 6.8 GHz (4.4 cm wavelength). To a distance limit of 250 pc, this encompasses the Taurus and Perseus star-forming regions. The final survey design was of 10 fields, centred on 10 primary targets, with $\sim30$ discs known from millimetre studies within the fields in total. The primary targets were selected from the catalogues of the Disks@EVLA and VANDAM surveys \citep{2014NMSymp, 2016ApJ...818...73T}.
%(Chandler et al. 2014; Tobin et al. 2016).
With 9 mm disc fluxes of 1--7 mJy, the primaries cover evolutionary stages from Class 0 protostars to Class II/HAe pre-main sequence stars. 

PEBBLeS has reached a survey sensitivity of $4-7$ $\mu$Jy beam$^{-1}$, with smallest beam sizes of approximately 60 mas. This is equivalent to a beam radius of $\sim3.5$ au for the nearest discs -- i.e. it separates the terrestrial-planet zone of solar system analogues from the orbits of the giant planets. Angular scales up to $\approx0\farcs7$ can be imaged.

Each observation was processed and calibrated using the e-MERLIN pipeline \citep[{\url{www.e-merlin.ac.uk}}; ][]{2021ascl.soft09006M}, with additional flagging where required and steps to ensure calibration consistency where several observations were taken within a few weeks. The targets were imaged with a range of weightings to investigate structure on different scales.
G26 also describe the procedure for fitting PEBBLeS disc-candidates (using {\sc uvmultifit}, \citealt{2014A&A...563A.136M}) and separating them from simultaneously-fitted jets in the $uv$-plane. This allows resolution of features smaller than the nominal restoring beam by a factor proportional to the signal to noise ratio \citep{2012A&A...541A.135M}.  We  first looked for a pair of Gaussian components, major axes within 30$^{\circ}$ of orthogonal (guided by ALMA and literature position angles), allowed to overlap. Parameters were adjusted until a good fit  was obtained with noise-like residuals (apart from some well-separated features probably  corresponding to resolved-out outflows). Components were added or subtracted as needed.  All ten sources had compact, central components with major axes within $1^{\circ} - 34^{\circ}$ of the ALMA disc and flux densities $>5\sigma_{\mathrm {rms\,noise}}$.  Some of these disc-candidates are as faint as 40 $\mu$Jy and as small as the orbit of Mars.

PEBBLeS is in the category of “thousand hour surveys”, when summing all the time used since the start of prototyping in 2014. It is expected that a similarly lengthy time might be dedicated to the study of demographic of pebble-forming discs with the SKA telescopes %\citep{2026aask.bookE..36G}.
\citep{Garufi01.2026.SKA}.

\section{Case studies}
\begin{figure}[h]
    \centering
    \includegraphics[width=0.49\columnwidth]{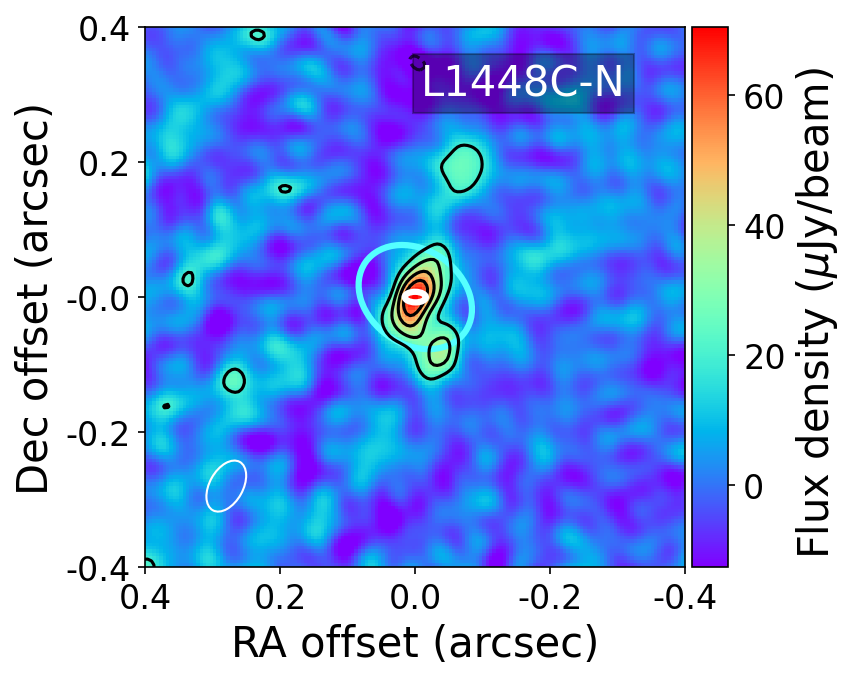} 
    	\includegraphics[width=0.5\columnwidth]{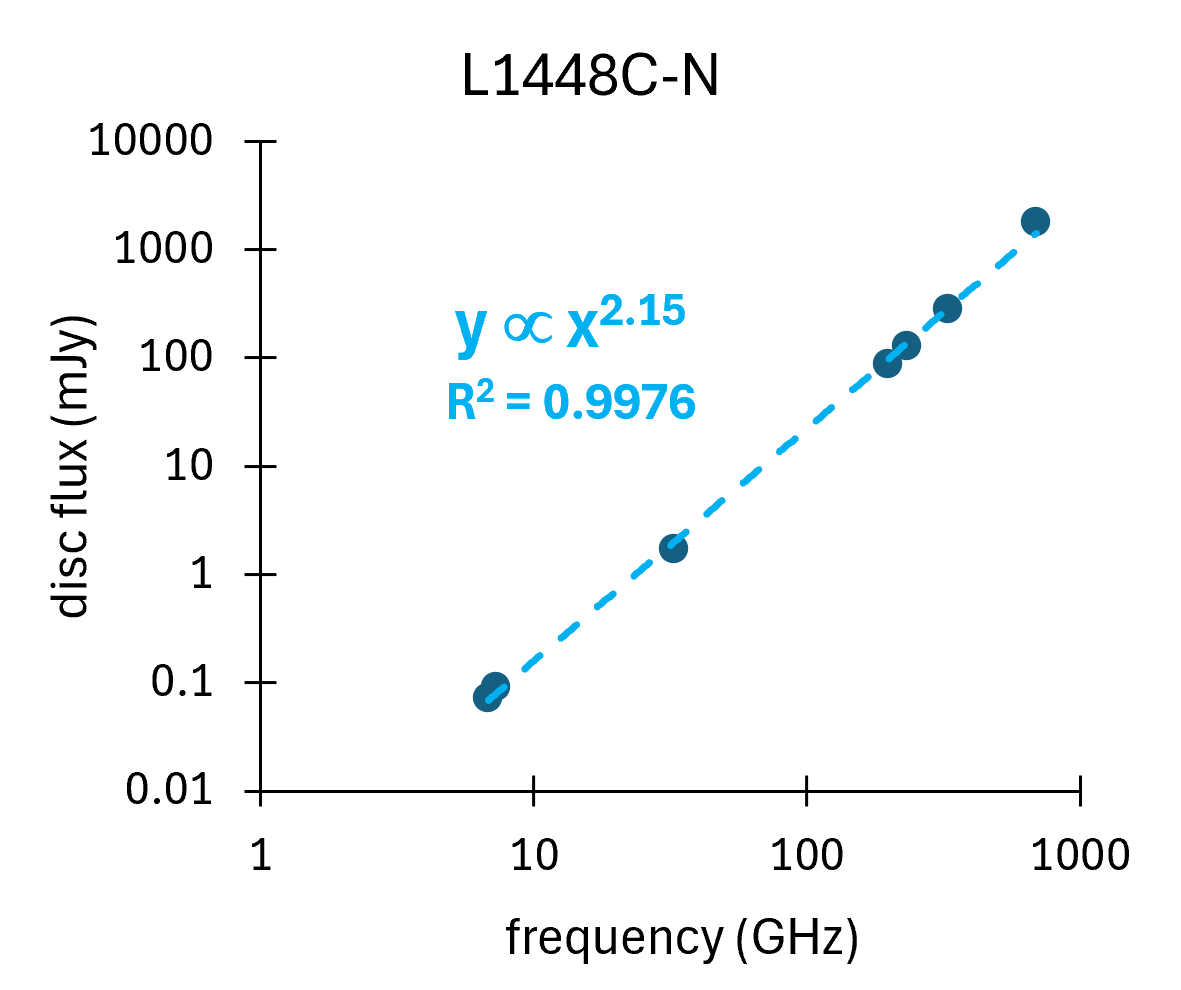}
        \caption{{\em Left}: L1448C-N as imaged by PEBBLeS.
          %The colour scale of the 6.8 GHz image is linear and
          The black contours are at 3, 5, 7, and 9 $\sigma_{\mathrm{rms}}$, where the  off-source noise  $\sigma_{\mathrm{rms}} \approx 6$ $\mu$Jy beam$^{-1}$ with a beam of $82\times50$ mas (white ellipse). A  disc (thick white ellipse) and jets were measured by $uv$-plane fitting. The small 5-$\sigma_{\mathrm{rms}}$ component to the SW is distinct from both disc and jets. The cyan ellipse represent a fit to an ALMA archival image at 233 GHz. %The sketched dotted curves illustrate possible locations of the outflow walls.
          {\em Right}: Spectral energy distribution of L1448C-N, using only data of comparably high angular resolution to PEBBLeS (central wavelengths and beam sizes in brackets). The points shown come from PEBBLeS, VANDAM, (4.1, 0.9 cm; beams 230, 65 mas)  
          \citep{2016ApJ...818...73T, 2018ApJS..238...19T}, 
          and the ALMA archive (1.5, 1.3, 0.9, 0.4 mm; beams 160, 150, 220, 150 mas), see Acknowledgements. The regression (blue dashes) has a coefficient $R^2 = 0.9976$ for a power-law index of 2.15.}
    \label{fig:L1448C-N}
\end{figure}

Figure~\ref{fig:L1448C-N} {\em Left} shows the e-MERLIN image of L1448C-N, a Class 0 protostar in Perseus. The background cyan ellipse shows the approximate size of the disc as seen by ALMA at 1.3 mm.
%and outlines (dotted) where low signal-to-noise structures at 4.4 cm 
The e-MERLIN data  $uv$-plane fit gave a compact 2D-Gaussian model containing 72 $\mu$Jy to FWHM (approximately east-west, thick white ellipse), forming a central disc. This represents 30\% of the total flux in fitted components.  The other components are more extended with peaks offset from the central disc in approximately orthogonal directions but overlapping the centre and  could be parts of the outflow \citep{2023MNRAS.522.1591T}. %(Toledano-Juárez et al. 2023).
The disc-candidate is oriented within $\sim$10$^{\circ}$ of the disc seen by VLA (9mm A+B configuration data released by VANDAM, {\url{https://dataverse.harvard.edu/dataverse/VANDAM}}) and the PEBBLeS feature has a radius of 4 au. The secondary peak seen at 4.4 cm (5$\sigma_{\mathrm{rms}}$ contour to the SW) does not coincide with any part of the jet system and lies within the ALMA disc area. 

Figure~\ref{fig:L1448C-N} {\em Right} shows the spectral energy distribution (SED) for L1448C-N using data between 6.7 -- 690 GHz at angular resolutions of 50 -- 230 mas, in order to reduce possible contributions from the outflows seen at lower resolution and in our additional fitted components. 
Jet emission at 4.4 cm (see Figure~\ref{fig:L1448C-N} {\em Left}) was separated spatially via the $uv$-plane separation process. This SED is well fitted by a continuous power law, $F_{\nu} \propto \nu^{2.15}$. The SW secondary peak is not included in the 6.8 GHz flux measurement point in this plot and this feature needs confirmation at other wavelengths.

\begin{figure}[t]
  \includegraphics[width=0.49\textwidth]{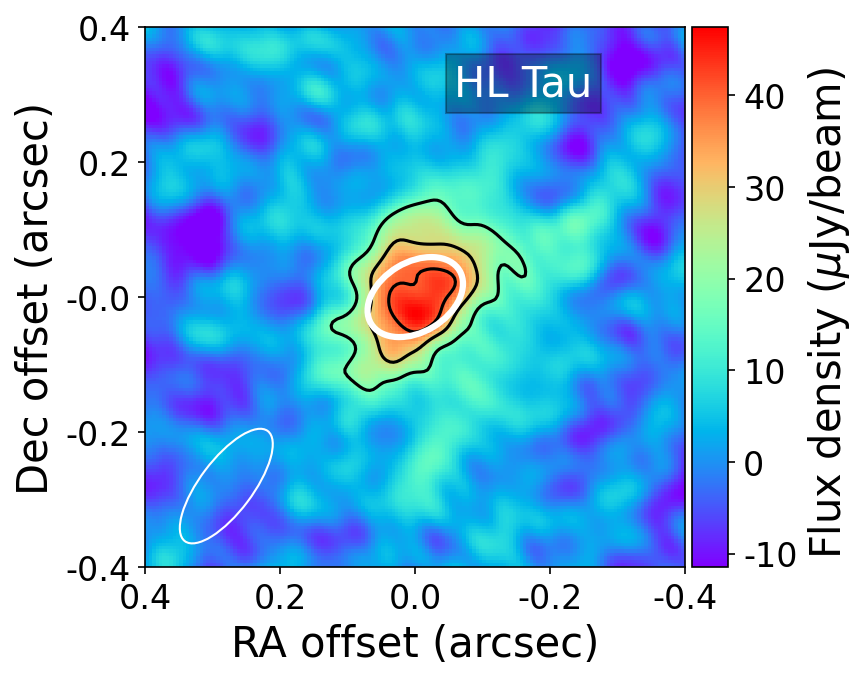}  
  \includegraphics[width=0.49\textwidth]{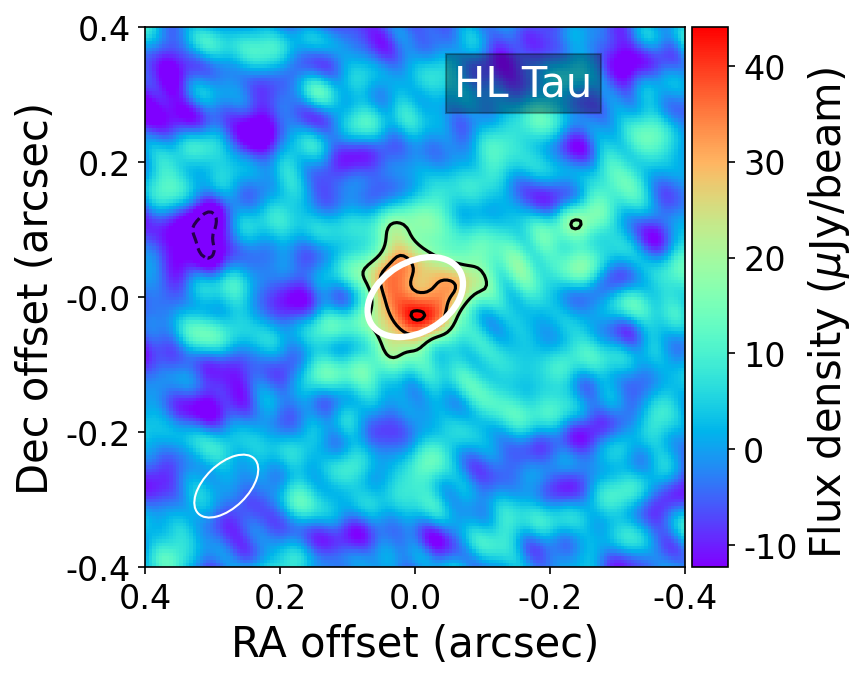} 
  \caption{e-MERLIN images of HL Tau at 4.4 cm, optimised for sensitivity ({\em left}), and for resolution ({\em right}). The black contours are at 3, 5 and 7 $\sigma_{\mathrm{rms}}$ where the off-source noise and restoring beams are 5.5 and 6.8 $\mu$Jy beam$^{-1}$ in (204 $\times$ 78) and (87 $\times$ 56) beams. The disc fitted in the $uv$-plane is overlaid (white outline).}
  \label{fig:HLres}
\end{figure}

Figure~\ref{fig:HLres} shows two 4.4 cm images of HL Tau, showing a well-resolved structure that is well aligned with the disc orientation as seen by ALMA \citep{2015ApJ...808L...3A}. The 
{\em left} panel is with a higher value of the robust cleaning parameter and the longer baselines are downweighted (tapered).  
%This gives a beam ($204\times78$) mas and $\sigma_{\mathrm{rms}} \sim6$ $\mu$Jy.   
The {\em right} image has a lower value of robust and weaker tapering, giving higher resolution.
%giving a beam ($87\times56$) mas and a higher $3\sigma_{\mathrm{rms}} \sim7$ $\mu$Jy.
Within 125 mas of the peak,  106 and 74 $\mu$Jy are recovered for the low and high resolution images, respectively  (compared with 102 $\mu$Jy from $uv$-plane fitting). The higher resolution image shows the ring-like distribution of the brightest and most compact emission,  not seen at shorter wavelengths. %No significant non-disc emission was detected in this system.
HL Tau has a large-scale outflow seen e.g. in H${\alpha}$ or CO observations, \citep{2012A&A...541A..16M, 2025A&A...704A.157B}.
\citet[][and references therein]{1996ApJ...470L.117W} analysed VLA observations at wavelengths between 0.7--6 cm, resolutions from 500--900 mas. A simple analysis of their cm-wave data predicts 460 $\mu$Jy flux at 4.4 cm within a comparable beam size.  However, they modelled the separate contributions of thermal and free-free emission. Extrapolating from their estimated thermal component at 1.3 cm using a spectral index of 2--2.6 predicts 4.4-cm emission of 120--60 $\mu$Jy, consistent with e-MERLIN results. This suggests that most of the flux detected at 90-mas resolution comes from a disc and the outflow is largely resolved-out. We were unable to fit additional components, although a minor fraction of free-free contamination cannot be excluded without more cm-wave observations at comparable resolution.  

\section{Advances with the SKA}

\citet{Garufi01.2026.SKA} 
%\citet{2026aask.bookE..36G}
discuss the powerful surveys of disc demographics that could be made with SKA-Mid. Here we consider observations analogous to PEBBLeS, but for the nearby star-formation regions of the southern sky. These regions are at comparable distances to the young stars we have observed in Taurus and Perseus. Hence we can estimate SKA-Mid performance by simulating an analogue of the well-studied disc of HL Tau. 
%but flipping its Declination from north to south.

SKA-Mid will benefit from large collecting area and wide bandwidth. In the funded delivery plan, AA$^{\ast}$ is expected to have 80 15-m antennas, and a 3.9 GHz instantaneous bandwidth in band 5a. We consider band 5a (centre frequency 6.55 GHz) as it is comparable to the 6.8 GHz effective frequency of PEBBLeS. The bandwidth with SKA-Mid is approximately eight times larger than the 0.512 GHz achievable with the e-MERLIN correlator. SKA-Mid also benefits from about a factor of two in collecting area over e-MERLIN (an exact value depends on antenna efficiencies). The beam of SKA-Mid will be smaller in band 5b, but massive discs may be opaque near the centre at these higher frequencies \citep{2019ApJ...877L..18Z}. %(Zhu et al. 2019).

\subsection{SKA-Mid simulation}

\begin{figure}[h]
    \centering
    \includegraphics[width=1.\columnwidth]{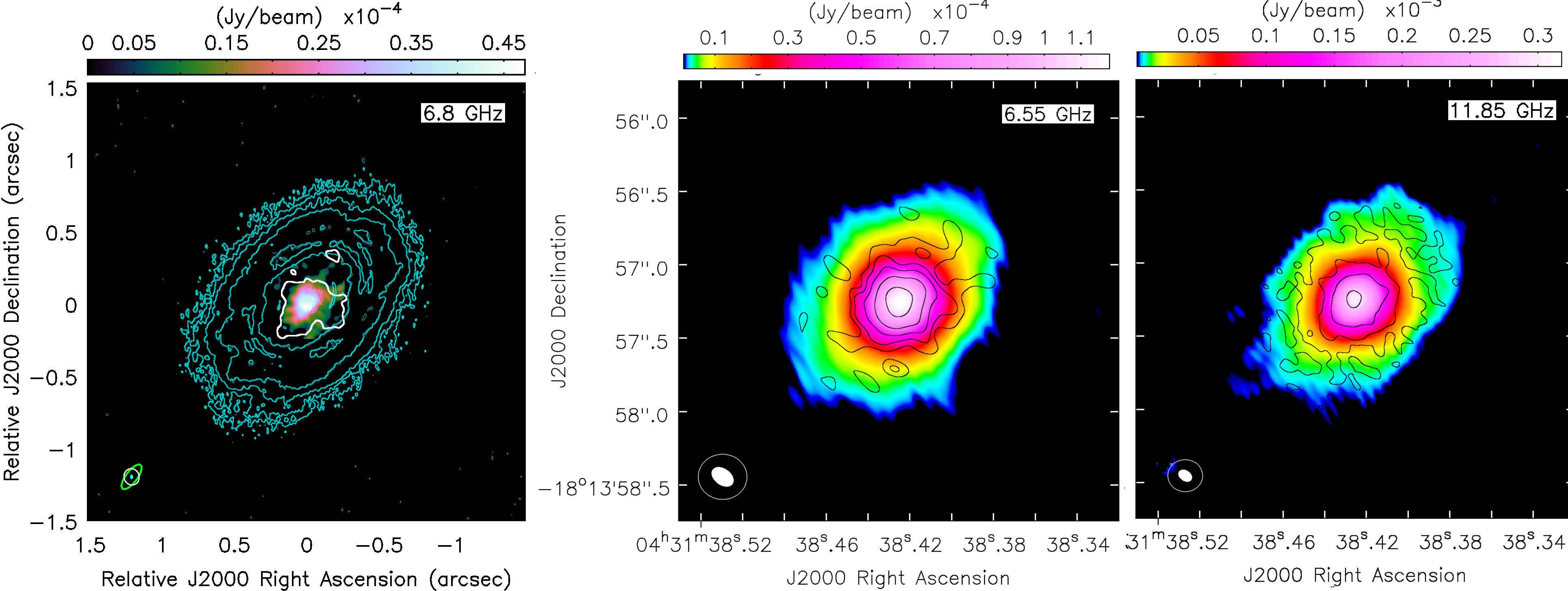}
\caption{{\em Left}: Overlay of original images of HL Tau: e-MERLIN at 6.8 GHz from PEBBLeS in colour scale with green beam (also shown in Fig.~\ref{fig:HLres}); VLA+Pie Town at 22 GHz \citep{2008MNRAS.391L..74G}, with white contour at $3\times16$ $\mu$Jy beam$^{-1}$ and white beam; ALMA archive data at 344 GHz \citep{2015ApJ...808L...3A} with cyan contours at $3,6,12... \times 24$ $\mu$Jy beam$^{-1}$ and small solid beam. These images were combined and used as models for AA SKA-Mid simulation with two weightings. {\em Middle}: simulation for SKA-Mid at 5a frequency, with colour scale blanked at $3 \times 0.5$ $\mu$Jy beam$^{-1}$ and contours at $3,6,12... \times 0.6$ $\mu$Jy beam$^{-1}$. {\em Right}: simulation at  5b frequency, with colour scale blanked at $3 \times 0.46$ $\mu$Jy beam$^{-1}$ and contours at $3,6,9,12... \times 0.7$ $\mu$Jy beam$^{-1}$.}
\label{fig:HLTau_sim}
\end{figure}

\subsubsection{Procedure}
The input images of HL Tau shown in Fig.~\ref{fig:HLTau_sim} {\em Left} were shifted to remove the proper motions between epochs (using the ALMA image as reference).  The VLA image had the central 10\% of emission removed to avoid the likelihood of detecting similar flux to ALMA in the inner disc -- this emission is probably optically thick at approximately $>10$ GHz, as the inner few tens of milliarcseconds show a dip in flux in the high-resolution e-MERLIN image, Fig.~\ref{fig:HLres} {\em right}. The flux densities were scaled to the SKA-Mid 5a and 5b frequencies using a spectral index of 2.5 for the ALMA data and 2 for the other observations.  The units in each case were converted to Jy pixel$^{-1}$, and other adjustments made for metadata consistency, and then the three images were summed. The Declination axis units were multiplied by --1 to emulate a Southern source.

The simulation goals were to investigate whether the SKA-Mid configuration AA$^{\ast}$ could detect and resolve (a) the 6.8 GHz inner emission seen by e-MERLIN; (b) the additional feature seen offset to the NW by the VLA at 22 GHz; and (c) lower-frequency emission from the more extended disc seen by ALMA. Not all these features of a real disc would necessarily contain significant quantities of pebbles, e.g. they might not have formed efficiently in the outer disc. 

We generated empty measurement sets representing SKA-Mid AA$^{\ast}$ configurations. We used 10 channels across the maximum continuum span of bands 5a and 5b, with 30 second integrations and a total time of 6 hr (for lower resolution images) and 24 hr (for higher resolution, to provide better visibility plane coverage).

We used the {\sc casa} \citep{2022PASP..134k4501C} {\tt sm} tool to add the model appropriate for each band, with noise estimated using the SKA sensitivity calculator \citep{2024SPIE13098E..2EB} extrapolated to a single channel and polarisation. At each band, we made
lower and higher resolution images, using robust values of 0 and --1.5 respectively. At high
resolution we used tapers of 0\farcs1 for 5a and 0\farcs06 for 5b.

\begin{figure}[h]
    \centering
	\includegraphics[width=0.9\columnwidth]{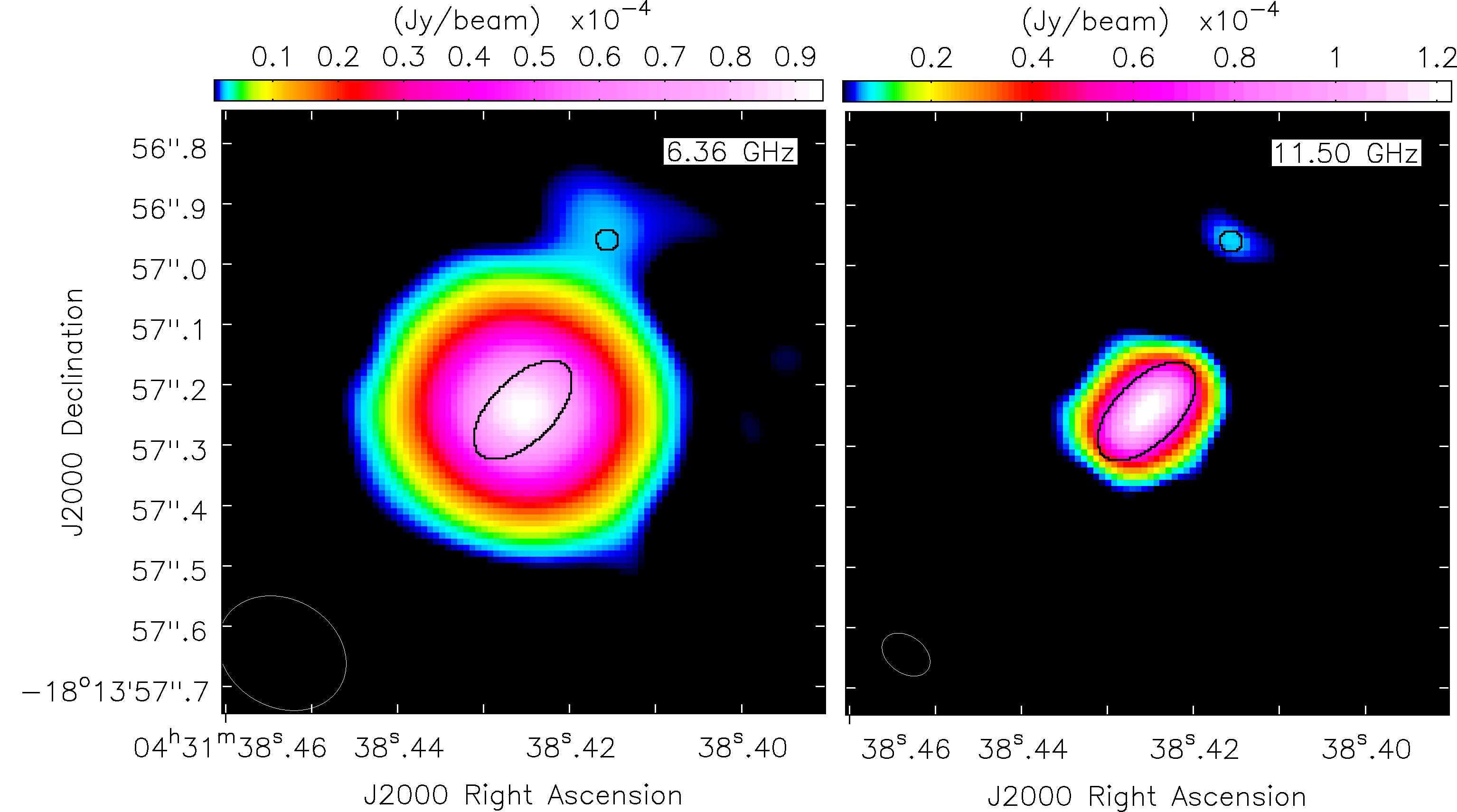}
    \caption{Simulation as for Fig.~\ref{fig:HLTau_sim}, but for a simple geometric model of a disc plus an additional clump of pebbles (black contours). The simulated images are shown in colour scale, with blanking levels set at multiples of the off-source noise. The left-hand plot is blanked at $3 \times 0.35$ $\mu$Jy beam$^{-1}$ and the right-hand plot is blanked at $3 \times 0.84$ $\mu$Jy beam$^{-1}$.}
    \label{fig:Modelsim}
\end{figure}

As not all discs will resemble HL Tau, we also tested a simple model comprising flat emission from a disc (axes of ($200\times100$) mas at position angle 135$^{\circ}$ E of N
) plus a clump  of 30-mas diameter, that could represent a compact region of dust aggregating around a planet. The values chosen are comparable to the HL Tau flux densities detected by e-MERLIN and the VLA, scaled using a spectral index of 2. For bands 5a and 5b  the small-clump flux densities are set to 3 $\mu$Jy and 10 $\mu$Jy respectively, with the extended disc at 40 times the clump’s flux density.  We generated measurement sets as above, and used the CASA sm tool and other tasks to simulate 24 hours of on-source observations, using configuration AA$^{\ast}$, with the results shown in Fig.~\ref{fig:Modelsim}.

\subsubsection{Outcomes}

Figure~\ref{fig:HLTau_sim} ({\em middle} and {\em right} panels) shows that, at both frequencies, the brightest emission is detected -- but details are not well-resolved, even at the highest resolution possible without excessive sidelobes / loss of surface brightness sensitivity. The candidate planet-related feature is detected, but it can only be distinguished by using prior information -- which would be hard to include if the orbital motion of such an object was unknown. The extended emission
detected by ALMA would be detected at lower resolution, if pebbles are present at such radii.

In the  simulation of a model system (Fig.~\ref{fig:Modelsim}), we find that at 11.5 GHz (band 5b) the clump is well-resolved, and it is recovered at 8 times the noise level at ($87\times61$) mas resolution. At 6.4 GHz (band 5a), the clump is detected at 8 times the noise at ($222\times179$) mas resolution, but if the weighting/tapering gives a beam smaller than approximately 150 mas the noise is too great, whilst at lower resolution the clump is merged with the disc. The main disc could be detected in a few hours, but the faint clump would require up to a couple of days at band 5a (including calibration and depending on Declination). Longer observations could be required for different pebble properties or if their aggregate surface area is less than used in these simulations. On the other hand, the wide bandwidth of SKA and the ease of lower-frequency observations very close in time will  facilitate identification and separation of non-thermal outflow emission.

We conclude that two different resolutions -- perhaps requiring two %pipelines to be run on
pipeline runs for the data -- could be needed for the most comprehensive disc science. 
Our experience shows that it could be necessary to decide optimum resolutions after  inspecting initial images, for discs with properties not well-known a-priori. For example, Fig. \ref{fig:HLTau_sim} shows that different weightings are needed to show the total disc flux density and to reveal the disc structure. $uv$-plane fitting is also a commonly-used technique for measuring disc parameters accurately, e.g. the {\sc frankenstein} or {\sc uvmultifit} packages, \citep{2020MNRAS.495.3209J, 2014A&A...563A.136M}. 
Further, the AA$^{\ast}$ configuration
has notable limitations for detecting small features related to planet formation. The high
resolution needed comes from baselines to only a single antenna, which is not ideal for imaging.
Working at higher frequencies will not necessarily solve the resolution problem, in the search for
planet-related structures in the inner disc, if massive discs are opaque above $\sim$10 GHz.

\section{Summary}

There are exciting prospects for large surveys of protoplanetary discs with SKA-Mid –- see Figure 4 in %\citet{2026aask.bookE..36G}.
\citet{Garufi01.2026.SKA}. 
In AA$^{\ast}$, a disc survey could already be much more time-efficient than PEBBLeS, or with similar times on-source an SKA-Mid survey could go around four times deeper per target than PEBBLeS has achieved. SKA-Mid could observe at even lower frequencies than band 5, where even the most massive discs would be optically thin near the centre, in the region where rocky planets may be forming. Longer SKA-Mid baselines are likely to be needed to resolve sub-structures related to such planet formation, for the discs of the southern sky.

Acknowledgements

We thank the wider PEBBLeS team and team members from Disks\@EVLA and VANDAM for all their help and inspiration, and Ian Heywood for help with simulations.
This paper makes use of the following ALMA data: 
2011.0.00015.SV, 2021.1.00508.S, 2018.1.01438.S. ALMA is a
partnership of ESO (representing its member states), NSF (USA) and
NINS (Japan), together with NRC (Canada), NSTC and ASIAA (Taiwan), and
KASI (Republic of Korea), in cooperation with the Republic of
Chile. The Joint ALMA Observatory is operated by ESO, auI/NRAO and
NAOJ.

\bibliographystyle{abbrvnat-maxbibnames4.bst}
\bibliography{PPPD} % if your bibtex file is called PPPD.bib

\end{document}